\documentclass[preprint2]{aastex63}

\usepackage{graphicx}
\usepackage[FIGTOPCAP]{subfigure}
\usepackage{amsmath,booktabs,threeparttable,url, bm}
\usepackage[hyphenbreaks]{breakurl}
\usepackage{CJKutf8}

\newcommand{\cntext}[1]{\begin{CJK*}{UTF8}{bsmi}#1\end{CJK*}}

\usepackage{natbib}
\received{\today}
\submitjournal{ApJ}

\shorttitle{ML for CCSN}
\shortauthors{Chao et al.}

\begin{document}

\title{Determine the Core Structure and Nuclear Equation of State of Rotating Core-Collapse Supernovae with Gravitational Waves by Convolutional Neural Networks}


\newcommand*{\NTHUCS}{Department of Computer Science, National Tsing Hua University, Hsinchu 30013, Taiwan}
\newcommand*{\NTHUP}{Department of Physics, National Tsing Hua University, Hsinchu 30013, Taiwan}
\newcommand*{\NTHUA}{Institute of Astronomy, National Tsing Hua University, Hsinchu 30013, Taiwan}
\newcommand*{\CICA}{Center for Informatics and Computation in Astronomy, National Tsing Hua University, Hsinchu 30013, Taiwan}

\newcommand*{\CTC}{Frontier Center for Theory and Computation, National Tsing Hua University, Hsinchu 30013, Taiwan}
\newcommand*{\NCTS}{Physics Division, National Center for Theoretical Sciences, Taipei 10617, Taiwan}
\newcommand*{\ISS}{Institute of Service Sciences, National Tsing Hua University, Hsinchu 30013, Taiwan}

\author{Yang-Sheng Chao (\cntext{趙仰生})}
\altaffiliation{These authors contributed equally to this work.}
\affiliation{\NTHUCS}

\author{Chen-Zhi Su (\cntext{蘇晨知})} 
\altaffiliation{These authors contributed equally to this work.}
\affiliation{\ISS} \affiliation{\CTC}

\author{Ting-Yuan Chen (\cntext{陳莛元})}
\affiliation{\NTHUA} \affiliation{\NTHUP} \affiliation{\CICA}

\author{Daw-Wei Wang (\cntext{王道維})}
\affiliation{\CTC} \affiliation{\NTHUP} \affiliation{\NCTS} 

\author[0000-0002-1473-9880]{Kuo-Chuan Pan (\cntext{潘國全})}
\affiliation{\CTC} \affiliation{\NTHUA} \affiliation{\NTHUP} \affiliation{\CICA}

\correspondingauthor{Kuo-Chuan Pan \& Daw-Wei Wang}
\email{kuochuan.pan@gapp.nthu.edu.tw; dwwang@phys.nthu.edu.tw}


\begin{abstract}

Detecting gravitational waves from a nearby core-collapse supernova would place meaningful constraints on the supernova engine and nuclear equation of state. 
Here we  use Convolutional Neural Network models to identify the core rotational rates, 
rotation length scales, and the nuclear equation of state (EoS), 
using the 1824 waveforms from \cite{2017PhRvD..95f3019R} for a 12 solar mass progenitor. 
High prediction accuracy for the classifications of the rotation length scales ($93\%$) 
and the rotational rates ($95\%$) can be achieved using the gravitational wave signals from $-10$~ms to $6$~ms core bounce. 
By including additional 48~ms signals during the prompt convection phase, 
we could achieve $96\%$ accuracy on the classification of four major EoS groups. 
Combining three models above, we could correctly predict the core rotational rates, 
rotation length scales, and the EoS at the same time with more than 85\% accuracy. 
Finally, applying a transfer learning method for additional 74 waveforms from FLASH simulations (\citealt{2018ApJ...857...13P}),  
we show that our model using Richers' waveforms could successfully predict the rotational rates from Pan’s waveforms even for a continuous value with a mean absolute errors of 0.32 rad s$^{-1}$ only. 
These results demonstrate a much broader parameter regimes our model can be applied for the identification of core-collapse supernova events through GW signals.
\end{abstract}

\section{INTRODUCTION}

Our knowledge of gravitational-wave (GW) astronomy has been advanced dramatically since the first detection of the binary black hole merger GW150914 \citep{2016PhRvL.116f1102A}. The newly released third Gravitational-wave Transient Catalog (GWTC-3; \citealt{2021arXiv211103606T}) enlarges our samples of gravitational events to 90 candidates, including binary black hole mergers, binary neutron star mergers \citep{2017PhRvL.119p1101A}, and neutron star-black hole coalescences \citep{2021ApJ...915L...5A}. 
One of the most promising following milestones will be the detection of gravitational waves from a nearby core-collapse supernova (CCSN) \citep{2021PhRvD.104j2002S}. CCSN are extremely energetic explosions ($E \sim 10^{51}$~erg) of massive stars at the end of their evolution and are likely the most promising multimessenger sources that could be detected with three different messengers, including multi-wavelength photons, multi-flavor neutrinos, and gravitational waves. The co-detection of multiple messengers enables us to comprehensively examine the underlying physics of the same energetic source at different times and in different regions associated with various physical processes.

Before an actual detection, one must understand the CCSN GW waveforms through prior theoretical and numerical calculations. However, unlike binary compact object mergers, our understanding of the CCSN waveform is significantly less mature, and the waveform calculations from multi-dimensional CCSN simulations are far more expensive than mergers. Until recently, thanks to the development of modern supercomputers, we are able to conduct a few high-fidelity, high-resolution 3D CCSN simulations that included sophisticated neutrino transport and state-of-the-art micro-physics \citep{2018ApJ...865...81O, 2018ApJ...861...10M, 2019MNRAS.487.1178P, 2019MNRAS.486.2238A, 2021ApJ...914..140P, 2022ApJ...924...38K} (also see \cite{2022arXiv220513438M} for a recent review). It is still, however, difficult to conduct CCSN simulations that cover the whole parameter domain, and the stochastic behavior of post-evolution makes the parameter estimation extremely difficult. 


\cite{2017PhRvD..95f3019R} provide 1824 axisymmetric general-relativistic CCSN simulations of a 12 solar mass progenitor with different rotational rates and nuclear equation of state (EoS), the largest set of gravitational waveforms update to date. However, the calculations are limited to the first 50 ms postbounce, 2D, and conducted with a single supernova progenitor. In their analysis and related works from \cite{2014PhRvD..90d4001A,2019ApJ...878...13P,2021ApJ...914...80P,2021PhRvD.103b3005A}, it is found that the information of the inner core angular momentum at bounce can be extracted from the GW bounce signal.  
Note that the rotational rates considered in \cite{2017PhRvD..95f3019R} spanned a wide range of angular rotational rates based on an empirical shellular rotational curve. 
Some high rotational rate models in \cite{2017PhRvD..95f3019R} are considered extremely rapid rotating and should be rare in nature, comparing to the rotating progenitor models from standard stellar evolution \citep{2000ApJ...528..368H, 2005ApJ...626..350H}. 
On the other hand, fast rotating models generate louder GW emissions than non-rotating models which make them easier to be detected with current GW detectors \citep{2019MNRAS.486.2238A, 2020MNRAS.493L.138S, 2020MNRAS.494.4665P, 2021ApJ...914..140P}.

\cite{2021PhRvD.103b4025E} uses a deep convolutional neural network (CNN) method to estimate the nuclear EoS of \cite{2017PhRvD..95f3019R}'s waveforms. 
They achieved 64-72\% correct classifications of the 18 nuclear EoSs. The correctness could be further improved to 91-97\% if they only consider the five EoS with the highest estimated probability. 
However, they did not provide any information on classifying the core rotational rates or core angular momentum, 
which are crucial physical parameters to examine the supernova engine.

\
\cite{2021PhRvD.103b3005A} use principal component analysis (PCA) and Bayesian parameter inference to extract the ratio of rotational to gravitational energy, $\beta= T/|W|$, from the same set of wavefroms in \cite{2017PhRvD..95f3019R}. 
For a galactic SN at a distance ($d=8.1$~kpc), they archived a $90\%$ credible interval of 0.004 of the $\beta$ for the Advanced LIGO. Similar accuracy could push to the Magellanic Clouds ($d\sim 50$~kpc) with the Cosmic Explorer \citep{2021PhRvD.103b3005A}. Although it is more natural to have slow rotating models, their analysis only focused on slow rotating models with $\beta < 0.07$.



In this paper, we adopt the same GW waveforms from \cite{2017PhRvD..95f3019R} and use a more powerful CNN model to identify not only the nuclear EoS but also the core angular rotational rates and the rotation length scales through a supervised learning method. The accuracy to identify these three parameters are $96\%$, $95\%$, and $93\%$ respectively, and reach $85\%$ to predict for all of these three parameters at the same time. Finally, we extend our CNN model through a transfer learning method for another dataset (see \cite{2018ApJ...857...13P}), and show that a very good prediction results could be still achieved with a much less amount of data even for a continuous core angular rotational rates. Our results demonstrate how a machine learning model could extract important CCSN parameters through the GW waveforms and has a great potential to be applied to a much wider parameter regimes.

The structure of this paper is organized as follows. In Section~\ref{sec_waveforms}, we describe the parameters of GW waveform we considered and the pre-processing method to re-sample the GW waveforms.  
In Section~\ref{sec_rotation}, we present the results of our CNN method to identify the core rotational rates and rotation length scale. The results of nuclear EoS classification is shown in Section~\ref{sec_eos}.
In Section \ref{Sec_TL}, We further show how our model could be extended to predict continuous core angular momentum through a transfer learning method with a new dataset. Finally, we summarize our results and conclude in Section~\ref{sec_summary}. 



%
%
%
%
\section{Gravitational Waveforms \label{sec_waveforms}}

Most of the gravitational waveforms used in this study are taken from the two-dimensional axisymmetric CCSN simulations with the {\tt CoCoNut} code \citep{2002A&A...393..523D} that are described in \cite{2017PhRvD..95f3019R}.
The waveform collection includes 1824 waveforms of a $12 M_\odot$ progenitor from \cite{2007PhR...442..269W} with 18 different nuclear EoS (see Table~I in \citealt{2017PhRvD..95f3019R}) and a shellular rotation profile,
\begin{equation}
\omega(r) = \omega_0 \left[ 1 + \left(\frac{r}{A}\right)^2 \right]^{-1}, 
\label{eq_rot}
\end{equation}
where $r$ is the spherical radius to the center of the star, $A$ is the rotation length scale to describe the degree of differential rotation, and $\omega_0$ is the initial rotational rate.
The provided GW signals start from $\sim -200$~ms to $50$~ms postbounce, where time zero represents the core bounce. The rotation length ranges from $A=300$~km to 10,000~km. We exclude 60 waveforms in this analysis due to the lack of core bounce in extreme rotating.
We also exclude additional 60 waveforms that have artificial enhanced or reduced electron capture rates ({\tt SFHo ecap0.1} and {\tt SFHo ecap10.0} in Table~IV in \citealt{2017PhRvD..95f3019R}) to avoid confusion with the SFHo EoS. 
One should note that the values of electron capture rates are not accurately established and the uncertainty of the electron capture rates might cascaded into  the core structure during collapse as reported in \cite{2017PhRvD..95f3019R} and \cite{2003PhRvL..90x1102L, 2022arXiv220209370J}. Therefore, it might affect the gravitational waveforms as well. However, the effects of electron capture rates are beyond the scope of this paper.  

\begin{figure}
	\includegraphics[height=2.3in]{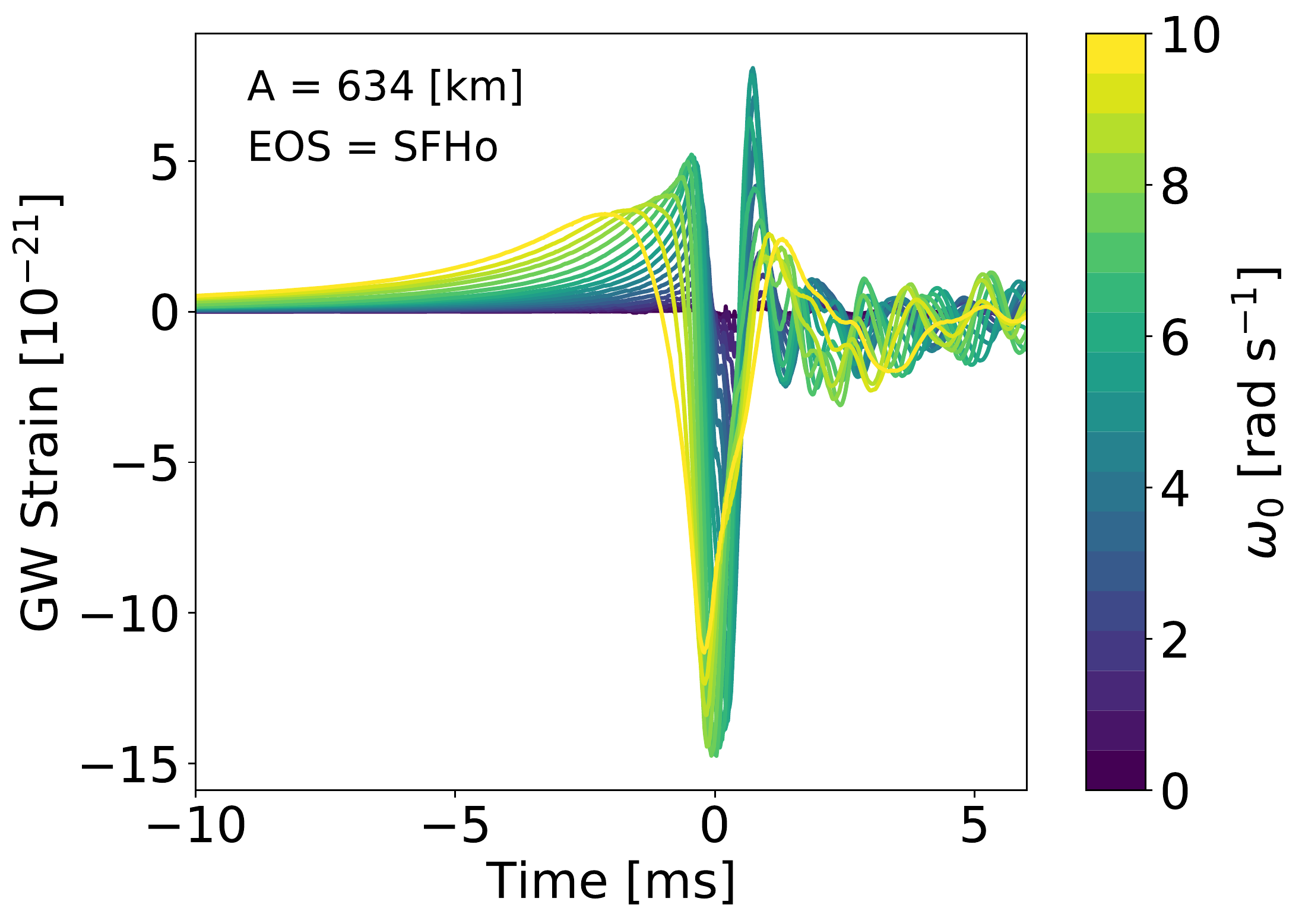}
	\includegraphics[height=2.3in]{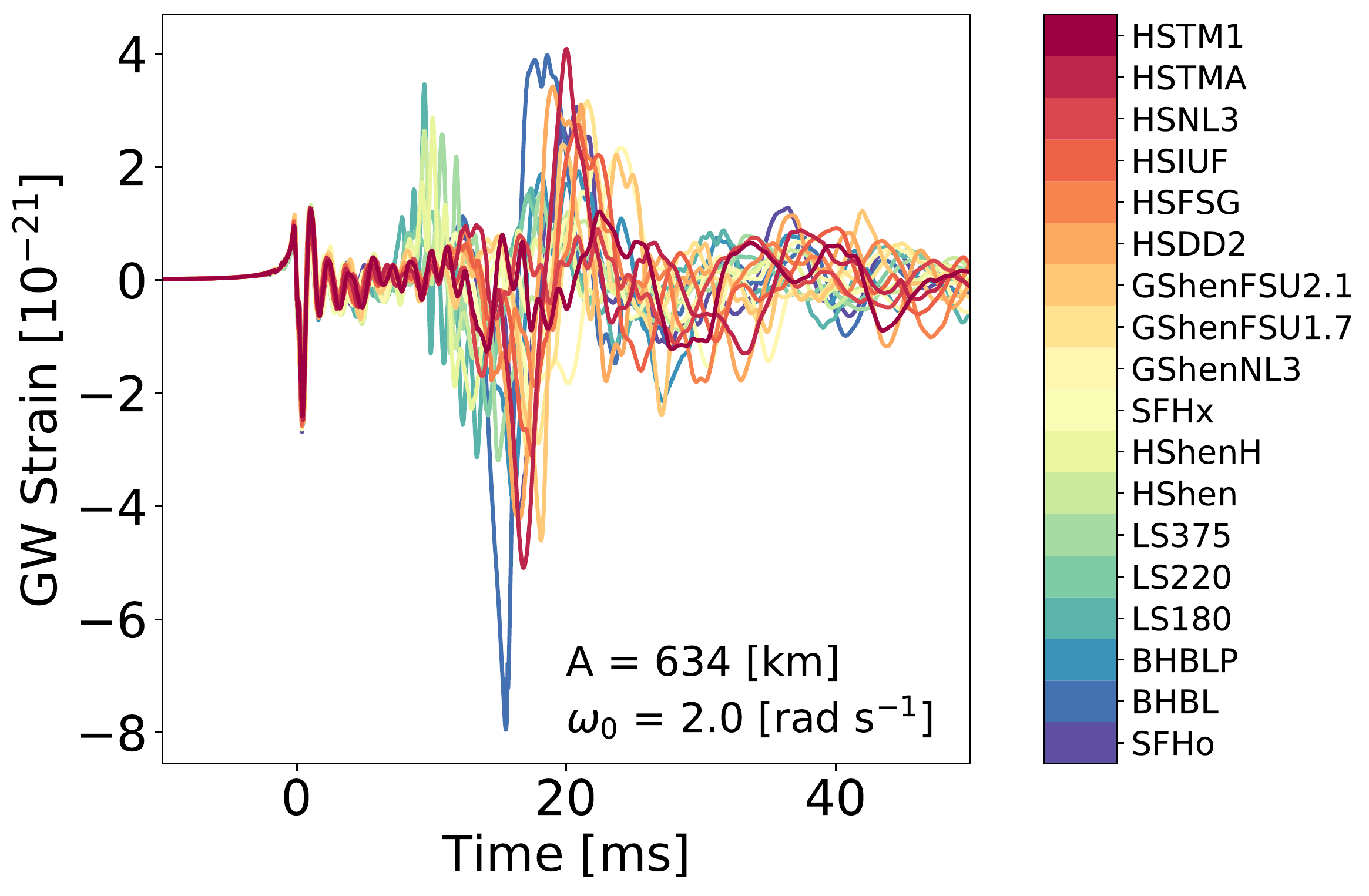}
	\caption{{\it Upper panel:}
	Gravitational wave strains around the core bounce ($t=0$~ms) with  $A=634$~km and SFHo EOS. Different colors represent different initial rotational speed $\omega_0$ (see Equation~\ref{eq_rot}). 
	{\it Lower panel:} Gravitational waveforms with $A=634$~km and $\omega_0=2.0$~rad~s$^{-1}$. Different colors indicate results obtained by different EOS. All waveforms are taken from \cite{2017PhRvD..95f3019R}.}
	\label{Fig:Typical GW signals}
\end{figure}

Figure~\ref{Fig:Typical GW signals} shows typical gravitational waveforms with different rotational speeds (upper panel) and different EoS (lower panel), taken from \cite{2017PhRvD..95f3019R}. Three clear features can be recognized as bellow.
First, the bounce signal around ($t=0$~ms) highly depends on the initial rotational rates ($\omega_0$) due to the rotational distortion from the centrifugal force (see the upper panel in Figure~\ref{Fig:Typical GW signals}). Second, the bounce signal has weak dependence in nuclear EoS \citep{2017PhRvD..95f3019R, 2021PhRvD.103b4025E} (see the lower panel in Figure~\ref{Fig:Typical GW signals}). 
Third, the GW from the prompt convection ($t \sim 20$~ms) depends on both the initial rotational rates and the nuclear EoS, but the dependence is not straightforward and the waveforms are stochastic. 
In this paper, we develop a CNN method to identify the rotational profiles ($\omega_0$ and $A$) and the nuclear EoS, assuming we could detect these waveforms from CCSN with high enough signal to noise ratios.   


\section{Identification of Core rotational Profiles }
\label{sec_rotation}


\subsection{Design of the CNN Model}
\label{Sec:CNN model}

In this paper, we use conventional CNN models to identify the rotational parameters, $A$ and $\omega_0$ in Equation~\ref{eq_rot}, and the associated equation of state (EoS), according to the GW signal simulated from various theoretical models in \cite{2017PhRvD..95f3019R}. CNN is a generalization of deep neural network, which was designed to mimic the architecture of human brain and is composed of one input layer and one output layer with multiple hidden layers in between. In each layer, there are a number of interconnected nodes (artificial neurons), which receives inputs from other neurons in previous layer and supply outputs to other neurons in the next layer. Each node performs a weighted sum computation on the values it receives from the input and then generates an output using a simple nonlinear function on the summation. CNN has some additional convolutional layers and pooling layers before the hidden layers, and therefore could capture the correlation between different input features for a better identification of images or signals.

To carry out the identification of these physical quantities, our input data are from 1704 simulated rotating core collapse signals \cite{2017PhRvD..95f3019R}, after taking away 60 cases without successful collapse and 60 cases with artificially enhanced or reduced electron capture rates. For the input feature, we first resample the GW signal to be a 2D ($x-y$) image with $256 \times 256$ pixels in a fixed time ($x\in(-10,6)$~ms) and strain ($y\in (-20 \times,20) \times 10^{-21}$) domain when focusing on the bounce signals. The time domain will be enlarged to $x\in(-10,54)$~ms if we want to include the information from the prompt convection phase.  


In order to provide better classification results, we take the experiences and advantages of computer vision and design our CNN model with four pairs of convolutional and max pooling layers before a dropout layer, which is followed by three fully connected layers before output. 
The number of output classes depends on the tasks to study (to identify $A$, $\omega_0$, or EoS) and will be explained in more detail later. The ratio of training data to the test data is 0.85:0.15, the learning rate is $10^{-5}$, the optimizer is chosen to be Adam, and the loss function is the standard categorical cross-entropy. All other hyper-parameters are listed in Appendix \ref{Appendix:model parameters} for details.

\subsection{Results to Predict the Rotational Length Scale, $A$}
\label{sec_core radius}

The first CNN model we develop is to identify the rotational length scale, $A$, from the observed GW signals. In the original data set, there are 5 values (i.e. $A=$300, 467, 654, 1268, and 10000 {km}) and hence the output layer of our CNN model has five nodes for the predicted probabilities. We do not distinguish the EoS in this model in order to have a larger sample size. The one of the largest probability is considered to be the predicted result and compared to the known labels in the test data. 

\begin{figure}
	\epsscale{0.95}
	\plotone{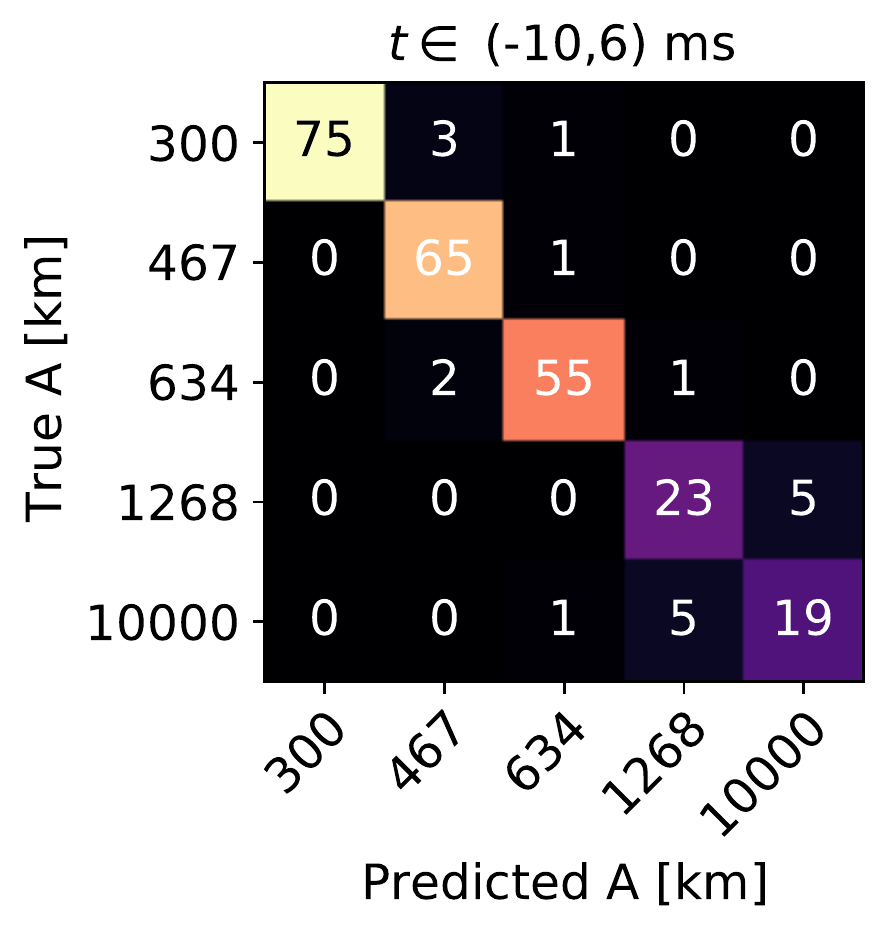}
	\plotone{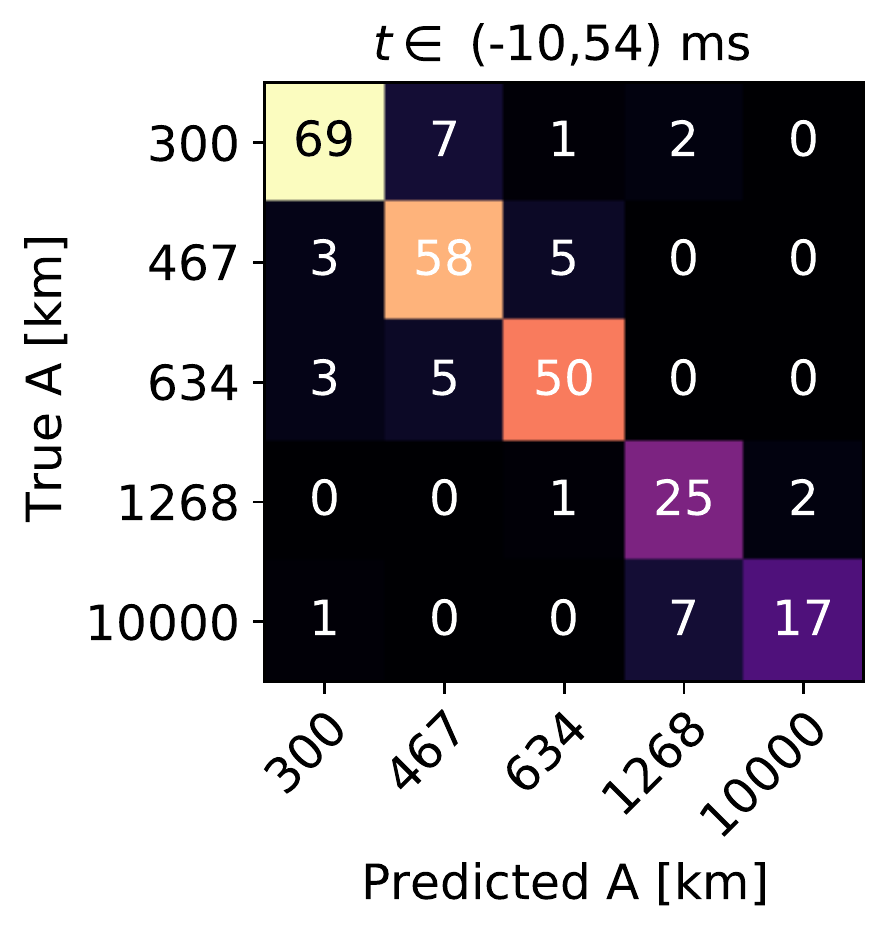}
	\caption{
	Confusion matrix for the prediction of rotational length scale, $A$. Five values of $A$ are included in the data set and two different time domains for the GW strain input are used for the model training and testing: the upper panel is for $t\in (-10,6)$~ms and the lower panel is for $t\in (-10,54)$~ms. The model accuracy is 93\% and 85\% respectively. Color codes are to highlight the significance of predicted data compared to the ground truth in the data set. 
	}
	\label{Fig:Confusion matrix of A}
\end{figure}

In Figure~\ref{Fig:Confusion matrix of A}, we show a typical result of calculated confusion matrices for the prediction of rotational length scale $A$. Results for using time period $(-10,6)$ and $(-10,54)$ are shown together for comparison. One could see that both results are very good and the overall accuracy are 93\% and 85\% respectively. We have to emphasize that the results obtained by short-time regime is significantly better than the results of longer-time regime, indicating that the most important features to identify the rotational length scale is the bounce gravitational wave signal. The chaotic behavior in the prompt convection phase ($\sim 20$~ms) will be difficult to train and is EoS depenent (see the lower panel of Figure~\ref{Fig:Typical GW signals}). 
This is consistent with one of the main conclusion in \cite{2017PhRvD..95f3019R}, which says that the GW bounce signal is insensitive to the EoS.


Besides of the overall performance of model prediction, we could further investigate how the model works for different values of $A$. One could see from the upper panel of Figure~\ref{Fig:Confusion matrix of A}, the correctness of prediction (ratio between the diagonal values and their neighboring off-diagonal values) for smaller $A$s (upper-left corner) is much better than that for larger $A$s (lower-right corner).
This result indicates that the GW signals should provide more important information at around core bounce
for the cases of small $A$ than the cases of larger $A$. 
Our results therefore not only demonstrate the importance of signal time for the prediction of the rotational length scale $A$, but also show how it depends on the value of $A$.

\subsection{Results to Predict Core Angular Rotational Speed, $\omega_0$}
\label{sec_angular momentum}

After predicting the rotational length scale with a very good accuracy, we further apply the same CNN model to predict the angular rotational speed, $\omega_0$, from the observed GW signals. In our dataset, there are 32 values of $\omega_0$, ranging from 0 to 15.5 rad/s with 0.5 rad/s step. Therefore, we will need to make a classifier for 32 types of results in the output layer of CNN. 

\begin{figure*}
	\epsscale{0.7}
	\plotone{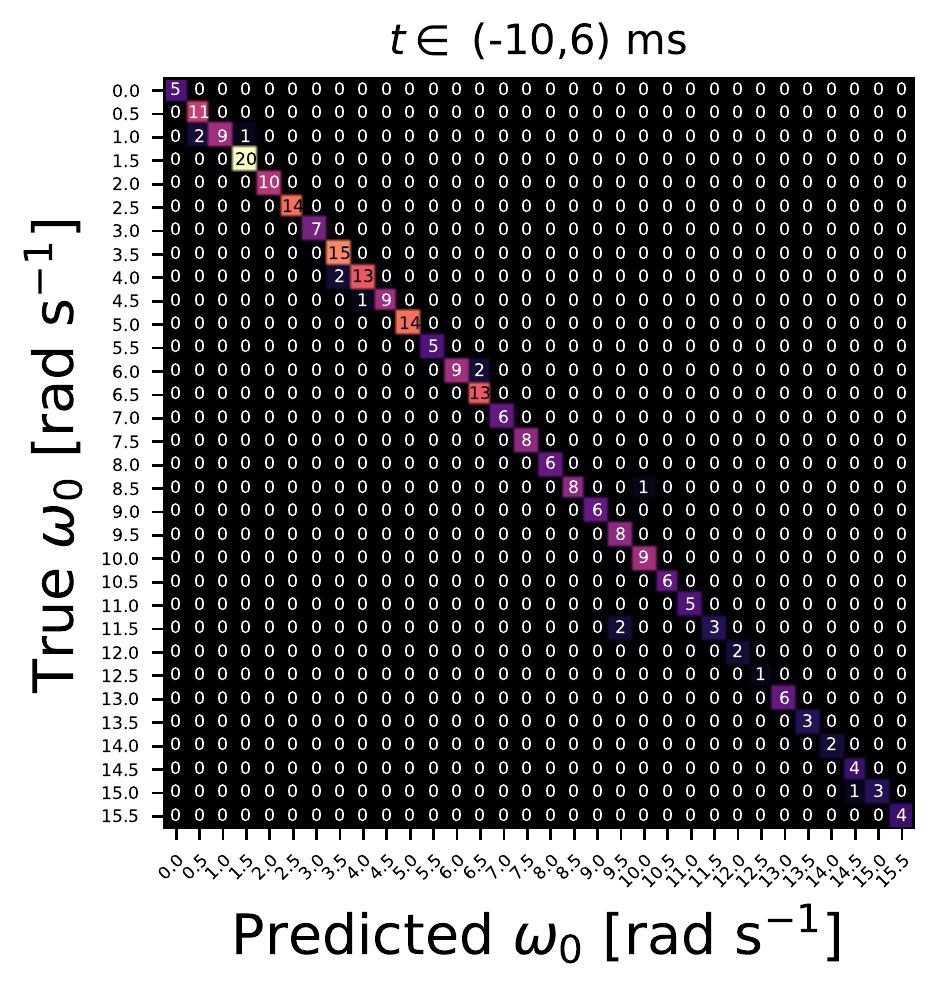}
	\plotone{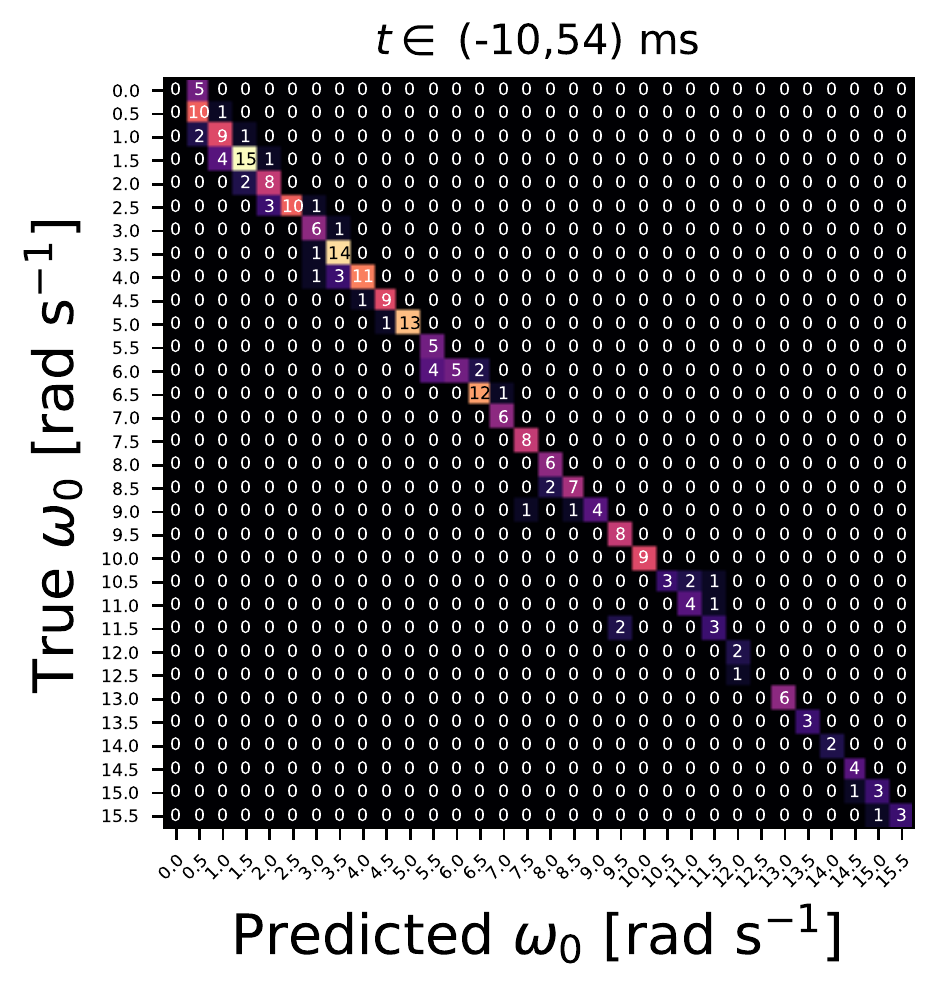}
	\caption{
	Confusion matrices for the prediction of core angular rotational speed, $\omega_0$. 32 values of $\omega_0$ are included in the dataset and two different time domains for the GW waveforms are used as shown in Figure~\ref{Fig:Confusion matrix of A}. The model accuracy is 95\% and 80\% respectively. Color codes are to highlight the significance of predicted data compared to the ground truth in the dataset. 
	}
	\label{Fig:Confusion matrix of omega}
\end{figure*}

Similar to the prediction of rotational length scale in the last section, here we train two CNN models to predict the rotational rate by using data with two GW signal time, $t\in (-10,6)$~ms and $t\in(-10,54)$~ms, respectively. Waveforms with different rotational length scale and EoS are all included for the training and test. In Figure~\ref{Fig:Confusion matrix of omega}, we show the corresponding confusion matrices for comparison. Their overall accuracy is 95\% and 80\% respectively, consistent with the prediction of rotational length scale in the last section. 
These suggest that the bounce GW signals consist more significant features than the prompt convection phase, and have little degeneracy with the rotational length scale (see the upper panel of Figure~\ref{Fig:Typical GW signals}), making the performance of machine learning much better if we only include the bounce signal. Another reason is that the GW in the prompt convection phase is very sensitive to the EoS and therefore might cause degeneracy while mixing waveforms from different EoS during the training.


\section{Identifying the Nuclear Equation of States, EoS }
\label{sec_eos}

Besides the identification of initial rotational parameters, one may expect a machine learning model to distinguish the 18 EoSs in the dataset, based on the GW waveforms they produced. However, different from initial conditions like $A$ and $\omega_0$, which lead to a pretty universal GW strain signals around the core bounce as described above, the effects of different EoS start to deviate from each other more significantly when the convection starts at around 20~ms postbounce (see the lower panel of Figure~\ref{Fig:Typical GW signals}). Therefore, in this section, we use the signals from core bounce and during the prompt convection phase ($t\in (-6,54)$) of the GW waveform as the input feature for the identification of EoSs. Naively, one might expect that the stochastic behavior during the prompt convection phase will not help to classify the EoS. However, information such as the starting time, strength, and buoyant frequency of the prompt convection are possible to be learned in neural networks.  

The naive application of our CNN model for such task does not have a good accuracy. 
Such a result is not surprising and can be understood from the following reason: different EoSs have different theoretical assumptions, approximations, or internal parameters for the nuclear matter. The effects of these hidden parameters could be very complicated in extreme conditions. It is therefore not reasonable to expect that our present dataset is large enough to distinguish these hidden parameters.

\begin{figure*}
	\epsscale{1.0}
    \plotone{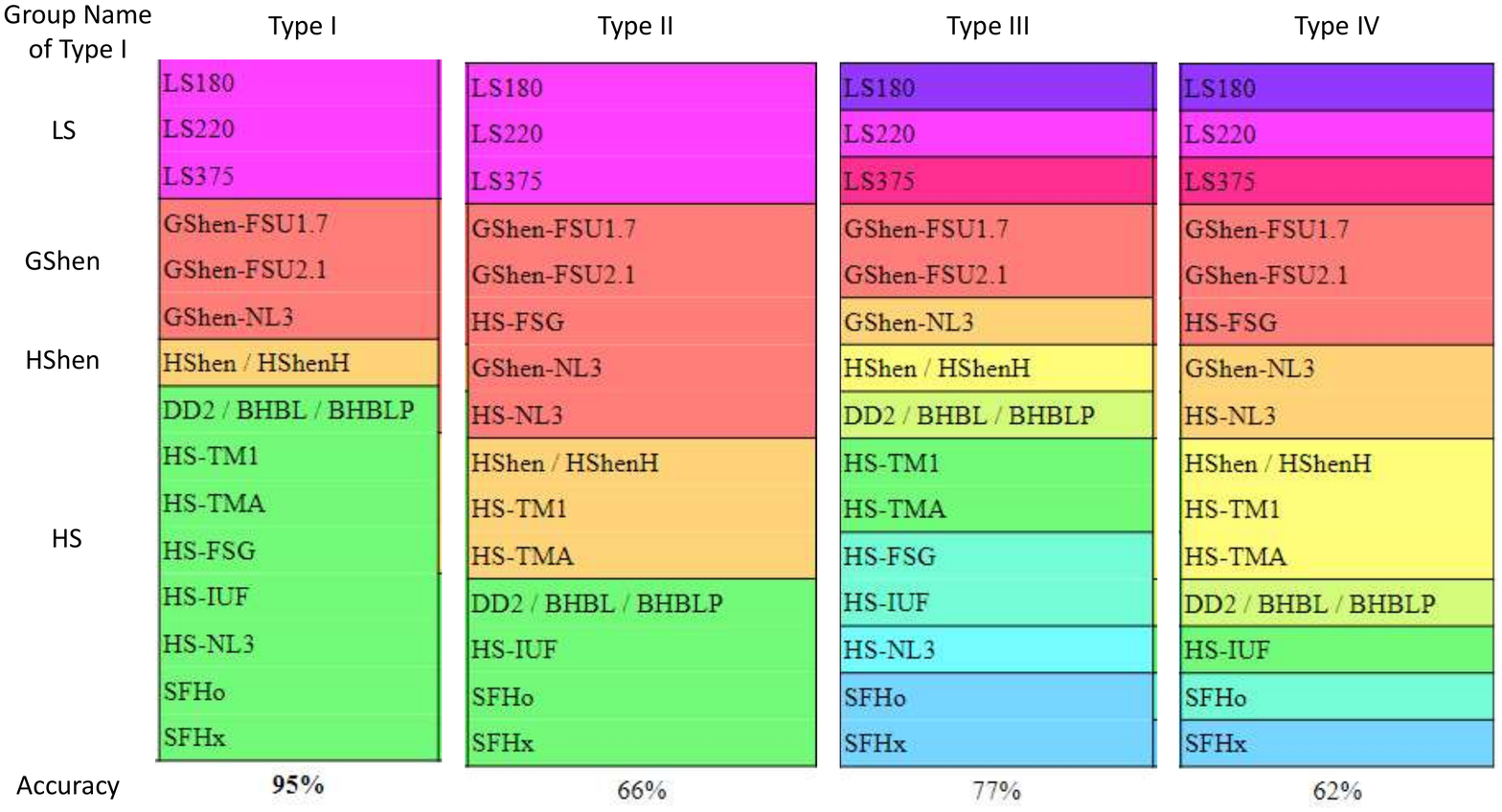}
	\caption{
	Four different grouping types of EoS. Each column indicates a grouping type, where EoS names with the same color are considered to be the same group. The predicted accuracy in the last line shows how well our CNN model could classify these groups for each type.}
	\label{Fig:grouping types of EoS}
\end{figure*}
\subsection{Clustering Equations of States}

In order to overcome such intrinsic challenges of supernova EoS estimation in the GW signals, we take the concept of unsupervised learning method in machine learning to cluster these 18 EoSs into several groups. Each group contain several EoSs, which are more "similar" to each other in some senses. We then expect that a machine learning model could be trained to capture the most important features for each group without confusion provided by other details of the EoSs inside each group. 

Since there are too many possibilities to cluster these 18 EoSs, here we just consider four types of clustering schemes, as listed in Figure~\ref{Fig:grouping types of EoS}, where each column is for one type. Within each type, the EoSs of the same color are considered to be the same group in our model training. The reason to consider these four types of grouping EoS is following:
In the first type of EoS clustering scheme, we simply group the EoSs based on the research group provided these EoS. For instance, the first group (LS), {\tt LS180, LS220} and {\tt LS375}, are provided from \cite{1991NuPhA.535..331L} using the compressible liquid-drop model for nuclei with different incompressibility parameters ($K$). The second group (GShen) includes three EoS ({\tt NL3, FSU1.7} and {\tt FSU2.1}) from \cite{2011PhRvC..83c5802S,2011PhRvC..83f5808S}. GShen group uses the relativistic mean field (RMF) calculations and with the RMF effective interaction NL3 or FSUGold. The {FSU2.1} version has an extra modification in high density. The third group (HShen) also uses RMF models but with TM1 parameters \citep{2011ApJS..197...20S}. The {\tt HShenH} EoS extra includes the effect from $\lambda$ hyperons, which could soften the EoS at high density. We do not distiguish {\tt HShen} and {\tt HShenH} becasue the hyperon effects is negligible at early core bounce in GW dataset. The forth group (HS) EoSs are derived from different RMF parameters (DD2/TM1/TMA/FSG/IUF/NL3) but has consistently transitions to nuclear statistical equilibrium (NSE) with thousands of nuclei at low density \citep{2010NuPhA.837..210H,2012ApJ...748...70H}. {\tt BHBL} and {\tt BHBLP} EoSs are based on the {\tt DD2} parameters but further include effects of $\Lambda$ hyperons and with or without repulsive hyperon-hyperon interactions \citep{2014ApJS..214...22B}. Using the Hempel's model, the SFHo and SFHx tuned the RMF parameters to fit the neutron star mass-radius observation \citep{2013ApJ...765L...5S}. 
The details of these 18 EoSs are also summarized in \cite{2017PhRvD..95f3019R}.

Since some of the GShen and HS EoSs use simliar RMF parameters, naively, we should group them together, assuming the same RMF parameters should give a similar EoS. Thus, in the second type of clustering scheme, we mix {\tt HS-FSG, HS-NL3} into the GShen EoS group and mix {\tt HS-TM1} and {\tt HS-TMA} into the HShen EoS group because of the similarity of RMF parameters.
The third and forth type of clustering schemes further divide the EoS groups into many subgroups based on the similarity of RMF parameters (see Figure~\ref{Fig:grouping types of EoS}). 

As for the third and fourth types of clustering scheme, we basically treat these 18 EoSs as independent groups, except for a few due to their similar internal parameters as described above.

\subsection{Classification Results: First Method}

We develop the same CNN model and train them for the classification of these different groups, i.e.~the EoSs in the same group are labelled the same in the output of our model. The calculated model accuracy for each type are shown in the last line of Figure~\ref{Fig:grouping types of EoS}. We can find that Type I with four groups are classified with 95\% accuracy, much better than the results obtained by other grouping types. It means that it is much easier for our CNN model to classify these four groups of EoSs in the high dimensional feature space. We may also say that EoSs within the same group are similar to each other, while EoS in different groups are relatively separated from each other, from the unsupervised machine learning point of view (although the results obtained here is by supervised machine learning method). 

\begin{figure}
	\epsscale{1.0}
	\plotone{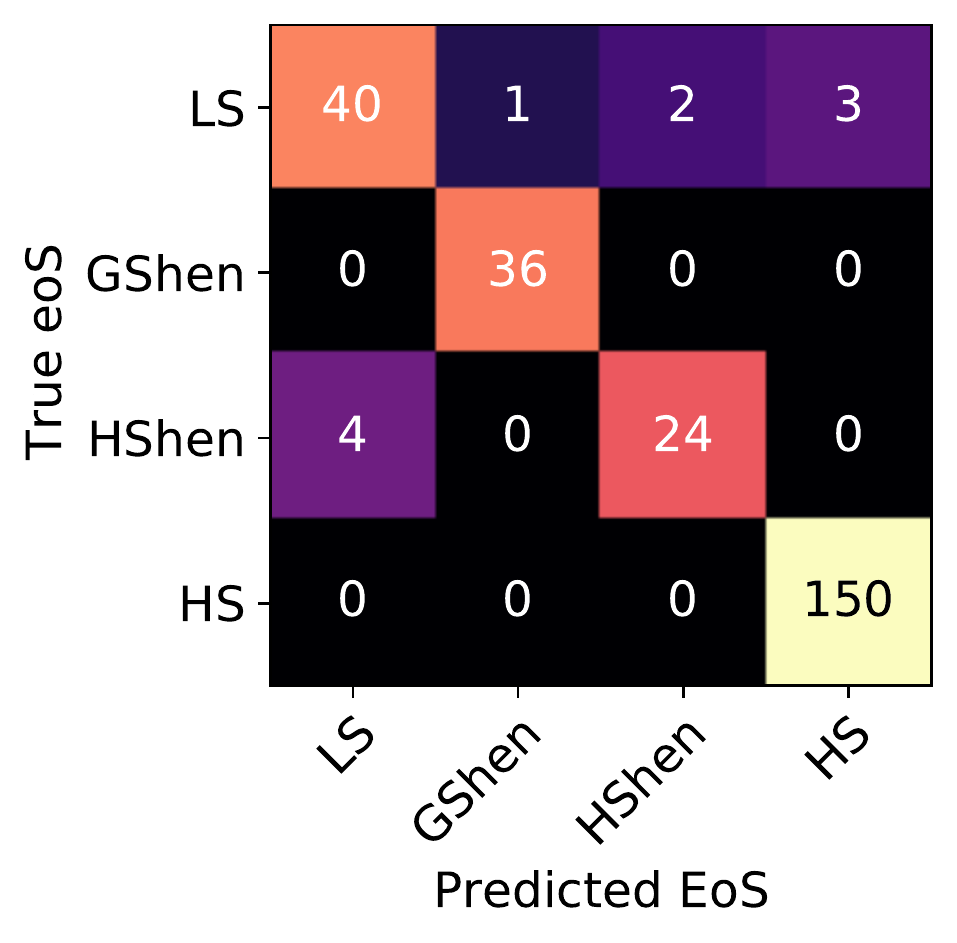}
	\caption{
	Confusion Matrix of the our CNN model to classify the four groups of Type I. For the convenience of discussion, the groups are labelled according to the name of most EoS inside, see Figure~\ref{Fig:grouping types of EoS}
	}
	\label{Fig:Confusion matrix of EoS}
\end{figure}

In Figure~\ref{Fig:Confusion matrix of EoS}, we further show the confusion matrix to classify these four groups for Type I. Note that, for the convenience of discussion, we have named these four groups according to their major EoS inside, see Figure~\ref{Fig:grouping types of EoS}. On could see that the overall performance of such classification is very impressive, because almost all EoSs of these cases are predicted correctly except for few cases in the off-diagonal elements.

The reason why the classification of Type-I group is better than other types is two folds: 
First, since there are less than 100 waveforms in each EoS, the sample size is not big enough to distinguish the all 18 EoS. Grouping into four groups could significantly improve the sample size for a better prediction. On the other hand, the fact that Type III grouping scheme (11 groups) shows better accuracy than Type II grouping scheme (4 groups) suggests that the sample size is not the only reason. Secondly, there are some hidden higher-dimensional features existing among different EoS providers, likely due to the physical assumptions used to derive these nuclear EoS.
Naively, we should expect that Type II grouping scheme should have a better accuracy due to closing RMF parameters, but accuracy drops from $95\%$ to $66\%$.
This is a little bit danger if we want to use future GW observations to constraint the nuclear EoS 
if the GW signals are sensitive to the underline physical assumptions/parameters. 
Although we could not exclude the possibility that our CNN model is biased by some numerical artificial features in the waveforms, the current results still strongly suggest that systematically investigation of these underlying assumptions/parameters could be necessary if we want to identify the correct EoS from the GW signals of supernova events.

\subsection{Classification Results: Second Method}

The above results for Type I clustering method could be also re-examined by another machine learning approach. We could train a binary classification model by using all data of EoSs in the same group to be the test data and data of all other EoSs to be the training data. Note that, different from the first method in the last section, the present model is to identify the corresponding core angular rotational speed, $\omega_0$, instead of the group label. (Similar results could also work to predict the corresponding rotational length scale, $A$.) The definition of the training data and the test data is to exam if the machine could learn their relationship, which should reflect if it is proper to divide these EoSs in such grouping scheme. 
We then repeat such model training by using data in different groups to be the test data for each time, and finally compare these results. 

In Table \ref{Table:Clustering re-examination I} we show the calculated results of model accuracy. One can see that the accuracy are not very high. Some of them are even less than 60\%, similar to random selection for such a binary classification. It is therefore reasonable to conclude that each group of Type I is  "far" from others in the high dimensional feature space. This is consistent with the confusion matrix results shown in Figure~\ref{Fig:Confusion matrix of EoS}, where the model is to identify the group labels directly, instead of identifying core angular momentum here. Note that, using this method to exam other types of Figure~\ref{Fig:grouping types of EoS} (not shown here) leads to a much higher prediction results between different groups, indicating certain correlation between these groups. 

In Appendix \ref{Appendix:EoS}, we further provide the second re-examination to check if there could be other subgroups within each group of Type I. The results shows that the grouping scheme of Type I is the most proper one for the identification of EoSs.

\begin{table}
\centering
\begin{tabular}{c c c c }
\hline\hline
Model & Test data & Training data & Accuracy \\
\hline
1 & LS & GShen+HShen+HS & 59\% \\
2 & GShen & LS+HShen+HS & 78\% \\
3 & HShen & LS+GShen+HS & 75\% \\
4 & HS & LS+GShen+HShen & 57\% \\
\hline
\hline
\end{tabular}
\caption{Model accuracy calculated for the first re-examination by using a given group to be the test data and others to be the training data (see the text). The poor binary classification results shows that these four groups in Type I clustering (see Figure~\ref{Fig:grouping types of EoS}) are reasonably well separated from each other, consistent with the classification results shown in Figure~\ref{Fig:Confusion matrix of EoS}.
}
\label{Table:Clustering re-examination I}
\end{table}
\subsection{Combined prediction of rotational length scale, core angular rotational speed, and Equation of State}

Since we have shown that we could use the same CNN structure to predict results of rotational length scale ($A$), core angular rotational speed ($\omega_0$), and EoS via using three different model parameters, it is instructive to check how the result could be if we predict these three parameters at the same time on the same GW waveforms. 

In Table \ref{Table:Prediction of all parameters}, we show the calculated accuracy for the combined prediction of these system parameters for the same test data, using models developed in previous sections. The obtained overall accuracy for these three parameters could be as high as 85\%, showing how reliable our CNN model could be applied to extract important physical parameters through the GW signals. 

\begin{table}
\centering
\begin{tabular}{cccccc}
\hline
\hline
 & $A \cap \omega_0 \cap$ EoS & $A \cap \omega_0$ & $A \cap $EoS & $\omega_0 \cap$ EoS\\
\hline
Accuracy & 85\% & 87\% & 89\% & 91\% \\
\hline
\hline
\end{tabular}
\caption{Accuracy for the prediction of rotational length scale ($A$), core angular rotational speed ($\omega_0$) and Equation of States (EoS) of the same test data in different combination of models.
}
\label{Table:Prediction of all parameters}
\end{table}
\section{Transfer Learning to Predict Continuous Core Rotational Speed \label{Sec_TL}}

In Section~\ref{sec_rotation}, we have shown the successful prediction of rotational length scale ($A$) and core angular rotational speed ($\omega_0$) by our CNN model. However, the training data obtained from  \cite{2017PhRvD..95f3019R} contains discrete values of $A$ and $\omega_0$ only, as described above. It is questionable how this model can be applied to predict results of continuous values, which may be more practical in realistic situation. Furthermore, it is also reasonable to expect that this model should be also applied for the data generated by other CCSN simulation code. After all, it may cost a lot of computational time for the generation of GW waveforms by the state-of-art multi-dimensional CCSN simulations with neutrino transport. 

In order to investigate how to extend our present model, trained by the data from \cite{2017PhRvD..95f3019R}, here we develop a new machine learning model for the prediction of a continuous core angular rotational rate, $\omega_0$, which are calculated by another simulation code, FLASH, from \cite{2018ApJ...857...13P} (not included in \cite{2017PhRvD..95f3019R}). This is based on the concept of transfer learning (TL) in modern machine learning methods.

\subsection{Traditional CNN Model for a Continuous Output}

The first naive method to apply our CNN model for a continuous core angular momentum is to use the same structure as before with the output layer replaced by a single neuron value, the corresponding rotational speed $\omega_0$. The input data is still taken from the GW waveforms for $t\in(-6,10)$ in \cite{2017PhRvD..95f3019R} (for all values of the rotational length scale, $A$), so that the CNN model is to simulate the function, $F_{CNN}(x(t))=\omega_0$, where the true value ($\omega_0$) is hidden implicitly inside the input GW signal, $x(t)$. We note that this is actually equivalent to a self-supervised learning approach, i.e. the output of neural networks is a physical quantity associated with the input data, instead of artificial labels in standard computer vision. 

However, since all the values of the core angular rotational speed in Richers' dataset are discrete, i.e. half-integer (HI) values in unit of rad s$^{-1}$ (see Figure~\ref{Fig:Confusion matrix of omega}), it is necessary to introduce additional data of continuous values, i.e. non-half-integer (NHI) for $\omega_0$, to test how well the above model works. In this work, we generate another dataset of GW waveforms, obtained by the FLASH simulations with the CCSN setup from \cite{2016ApJ...817...72P,2018ApJ...857...13P}. 
Pan's waveforms are derived from two-dimensional hydrodynamics simulations with the Isotropic Diffusion Source Approximation (IDSA; \citealt{2009ApJ...698.1174L}) for the neutrino transport, and with an approximated general relativistic solver based on the description in \cite{2006A&A...445..273M}. We use the same progenitor model s12 and the SFHo EoS for consistency. We also use the same initial rotational curve as defined in Equation~\ref{eq_rot}, but adopt both half-integer and non-half-integer $\omega_0$. The total Pan's waveforms include 52 half-integer $\omega_0$ and 22 non-half-integer $\omega_0$ in the range $\omega_0\in(0,6.0)$ for different values of $A$s. In the calculation below, we use half of them for the test data (the other half will be used for the transfer learning, see below).

\subsection{Transfer Learning Model for a Continuous Output}

Another approach to predict the continuous core rotational parameters is to apply the transfer learning (TL) method \citep{2016TL}, which is a two-step training process: First, we use the original (Richers') data to "pre-train" a CNN model for the prediction of a continuous $\omega_0$, same as the process mentioned above. Secondly, we fixed the model parameters in the CNN layers and let only the parameters of the last layer of neurons to be modified by new data, which is obtained by Pan's waveforms. These two-step TL makes the internal parameters of our CNN model to have a good initial values, but to be fine-tuned by the new data for the final purpose of simulations. 

We note that different from the traditional CNN model, TL takes the advantage of the original abundant data (here is Richers') for the pre-training process, but modifies only a small portion of its internal parameters of the CNN model by the new data to achieve a better performance. As a result, the second step training could become very efficient compared to the traditional method. As an example, here we use half of Pan's waveforms to be the training data in the second stage and use the other half for the test, which consists of 
37 GW waveforms with 26 HI $\omega_0$s and 11 NHI $\omega_0$s, randomly selected from different values of $A$ and $\omega_0$. It is expected to be a promising method for the extension of our model to other new types of data, even generated by other supernova code and/or with continuous values.

\subsection{Comparison between Models with and without TL} 

\begin{figure}
	\epsscale{1.2}
    \plotone{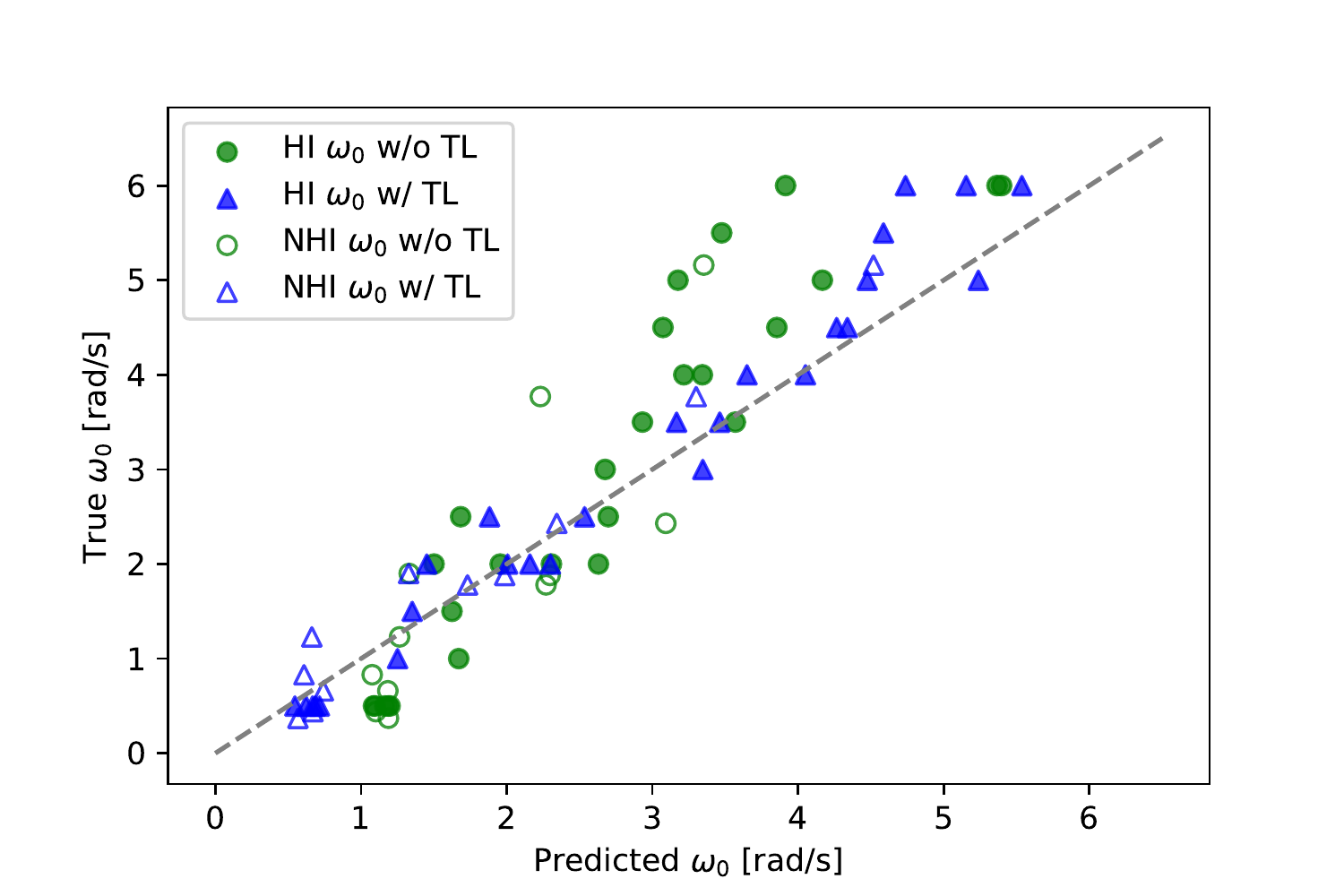}
	\caption{
	The predicted values of $\omega_0$ using two different models: our CNN model without TL (green circles) and a TL model trained by using additional waveforms from Pan (blue triangles). Filled symbols represent results for HI $\omega_0$ and opened symbols indicate results for NHI $\omega_0$. The diagonal dashed line indicates the correct results.
	}
	\label{Fig:TL for omega}
\end{figure}

In Figure~\ref{Fig:TL for omega}, we show calculated results for both models (with and without transfer learning). 26 of these test data, generated from Pan's waveforms (with different rotational length scale, $A$), are HI (filled circles) and 11 of them are NHI (open circles). One can see that for the results calculated by model without TL (green filled/open circles), the predicted values are in general underestimated when compared to the ground truth. The deviation might be resulted from different gravity treatments between two codes, where \cite{2017PhRvD..95f3019R} use the conformal flatness condition (CFC) approximation but \cite{2018ApJ...857...13P} use an effective general-relativistic potential to handle gravity. This shows that the naive application of our CNN model to a continuous value of $\omega_0$ or to results obtained by other supernova code cannot provide reliable prediction.

On the other hand, when the TL approach is included as shown by blue filled/open triangles (for HI/NHI values of $\omega_0$ respectively), the predicted values are in general much closer to the expected value (straight dashed line) for almost all values of $\omega_0$. We note that even for the non-half-integer values (open triangles), the predicted results are also much better than the traditional model without TL. These results shows that we could apply transfer learning method to successfully extend our model to a more general situation, even if continuous values of $\omega_0$ and/or different EoS are considered.

To quantify the different results of these two models, we calculated the mean absolute error (MAE) for the prediction of continuous values from their corresponding truth values, i.e. MAE$=\frac{1}{N}\sum_i^N|y_i-y_i^\ast|$, where $y_i$ and $y_i^\ast$ are the predicted values and the true values of $\omega_0$ respectively. $N$ is the total number of test data. We find that MAE$=0.72$ rad~s$^{-1}$ for the model without TL, and becomes $0.32$ rad~s$^{-1}$ only after including TL. Note that, this result is obtained by using only 37 GW waveforms from Pan's waveforms for the second step training. This shows that we could apply TL for the prediction of a more general situation: continuous core angular momentum even calculated by other simulation codes. 

In summary, our CNN model achieves $95\%$ accuracy on evaluating $\omega_0$ with a $0.5$~rad~s$^{-1}$ interval with Richers' waveforms. After applying a TL method, we could recognize continue $\omega_0$ with a MAE$=0.32$~rad~s$^{-1}$ with Pan's waveforms. 
In comparison with the PCA method described in \cite{2021PhRvD.103b3005A}, 
Alfe \& Brown achieved a $90\%$ credible interval of 0.004 of the $\beta$ for a SN from the galactic center and observed by the Advanced LIGO.
Although the value of $\beta$ depends on the core structure of a collapsar and cannot directly translated into $\omega_0$, the credible interval of $0.004 \propto \frac{1}{2} I \Delta \omega_0^2/|W| \propto \Delta \omega_0^2$, leading to $ \Delta \omega_0 \sim 0.2$~rad~s$^{-1}$, which is comparable with our CCN network for slow rotating models. 
It is known that the statistical methods (e.g. \citealt{2021PhRvD.103b3005A}) are based on certain hypothesis of certain models and calculate the correlation between observed signals and these model parameters. The CNN model in this work is a purely data-driven approach and therefore could be applied in a more general situation. Furthermore, the waveforms considered in this work also include a wider range of rotational rates, but effects due to GW background noise or glitches are not considered yet in this work.

%
%
\section{SUMMARY AND CONCLUSIONS \label{sec_summary}}

We have trained two CNN Models to identify the core rotational parameters 
(i.e. $A$ and $\omega_0$ in Equation~\ref{eq_rot}) 
for the s12 progenitor from \cite{2007PhR...442..269W}, 
using the 1824 gravitational waveforms in \cite{2017PhRvD..95f3019R}. 
We find that the classifications of the these rotational parameters are more accurate 
when using only the GW signals around the core bounce (-10~ms $< t <$ 6~ms). 
We have achieved $93\%$ and $95\%$ accuracy on identifying the core rotational length scale ($A$) 
and core rotational rate ($\omega_0$), respectively. 
These results suggest that the bounce GW signal is an excellent diagnosis for determining the core angular momentum of the collapsing progenitor and less sensitive with the nuclear EoS. 
This is also consistent with the previous finding in \cite{2008PhRvD..78f4056D, 2010A&A...514A..51S, 2014PhRvD..90d4001A, 2017PhRvD..95f3019R, 2021ApJ...914...80P}, but first time to be confirmed with neutral networks.  

We further trained the third CNN model to classify the types of nuclear EoS (see Figure~\ref{Fig:grouping types of EoS}) with the same set of gravitational waveforms but include extra 48~ms signals during the prompt convection phase (-10~ms $< t <$ 54~ms). 
It is found that the Type~I grouping scheme in Figure~\ref{Fig:grouping types of EoS} shows the best 95\% accuracy. 
Type~II grouping scheme is expected to have a better accuracy since they share similar RMF parameters but mixing EoSs provided by different research teams. 
However, the results of Type~II grouping scheme show significant less accurate results (66\%) with similar sample size as Type I grouping scheme, 
suggesting that either there are hidden hyper-dimensional features from the physical assumptions that used in these EoS research teams, or our CNN models learned other numerical artificial features in these EoS tables. If the former one is correct, one should be more careful when using these waveforms to constraint the EoS in future GW searches. 

Combing the above three CNN models, we have achieved an overall 85\% accuracy to predict $A$, $\omega_0$, and the EoS group at the same time. 
To further extend our trained model, we also have applied a transfer learning model that uses additional 74 waveforms from the FLASH-IDSA code used in \cite{2018ApJ...857...13P}.
By just using 37 waveforms during the training process, we could successfully predict the rotational rates $\omega_0$ in Pan's waveforms with a $MAE = 0.32$~rad~s$^{-1}$, even for non-half-integer $\omega_0$.
The transfer learning model we used opens the possibility to train CNN models with handful 3D waveforms from state-of-the-art 3D CCSN simulations that are extremely computationally expensive.  
With this technique, our trained model can be further extended or calibrated with more realistic waveforms from new simulations or future observations. 

However, one should also note that all the waveforms used in this paper are with the s12 progenitor from \cite{2007PhR...442..269W} and the duration of the GW signals are limited within the first 50~ms postbounce. 
In realistic, different progenitor mass has a different compactness at collapse and therefore should give different GW signals. In addition, late-time GW signals are crucial to understand the evolution of the proto-neutron star and the high-density nuclear theory when closing to BH formation \citep{2018ApJ...857...13P, 2021ApJ...914..140P}. Furthermore, 3D fluid instabilities \citep{2018ApJ...865...81O, 2018ApJ...861...10M, 2019MNRAS.487.1178P, 2019MNRAS.486.2238A, 2021ApJ...914..140P, 2022ApJ...924...38K} or magnetic filed effects \citep{2014ApJ...785L..29M, 2021MNRAS.503.4942O, 2021ApJ...906..128K} are not developed yet or ignored in the considered time duration of this study. 
Moreover, GW noise and glitches in realistic waveforms should reduce the accuracy of parameter estimation in our CNN models. Improving the training with noises or including noise cleaning techniques 
\citep{2019PhRvD..99d2001D, 2020PhRvR...2c3066O} are crucial for future GW detection.
Even though, our results provide positive hints for future CCSN parameter estimation using machine learning techniques.   

\begin{acknowledgments}

This work is supported by the Center for Informatics and Computation in Astronomy (CICA), National Center for Theoretical Sciences, and by the Higher Education Sprout Project funded by the Ministry of Science and Technology and Ministry of Education in Taiwan. KCP is supported under the grant MOST 110-2112-M-007-019, 111-2112-M-007-037, and DWW is supported under the grant MOST 110-2112-M-007-036-MY3.
The FLASH simulations and data analysis have been carried out on the CICA Cluster at National Tsing Hua University.

\software{FLASH \citep{2000ApJS..131..273F, 2008PhST..132a4046D}, yt \citep{2011ApJS..192....9T}, Matplotlib \citep{2007CSE.....9...90H}, NumPy \citep{2011CSE....13b..22V}, SciPy \citep{2019zndo...3533894V}, Scikit-Learn \citep{2011SKLearn}, Keras \citep{2015keras}, TensorFlow \citep{2015TenserFlow}}

\end{acknowledgments}


\appendix
\section{Hyper-parameters of the CNN Model}
\label{Appendix:model parameters}
In this Section, we summarize our CNN architecture and hyper-parameters used for the machine learning calculation in Tabel~\ref{tab_cnn1}. The structure is similar to the standard CNN model with several convolutional/pooling layers, followed by deep neural networks with fully connected layers. All the activation functions are ReLu for these hidden layers. Compared to \cite{2021PhRvD.103b4025E}, we structure has more convolutional/pooling layers as well as deeper fully connected layers for a better simulation results. We also introduce dropout layers to reduce the possible over-fitting. The total number of internal parameters is similar to \cite{2021PhRvD.103b4025E}.

Moreover, in order to apply this structure to various models described in the paper, we fix the hyper-parameters of this structure and allow only the number of output neurons in the last layer to be modified. For example, it is 5 for the prediction of rotational length scale, $A$ (see Fig. \ref{Fig:Confusion matrix of A}), 32 for the prediction of core angular rotational speed, $\omega_0$ (see Figure~\ref{Fig:Confusion matrix of omega}), is 4 for the classification of EoS in Type I group (see Figure~\ref{Fig:Confusion matrix of EoS}), and is 1 for the prediction of continuous core angular rotational speed (see Fig. \ref{Fig:TL for omega}. Our numerical results of training/test shown in the text has demonstrate that such a structure has a pretty good balance between the generality for different tasks and the simplicity for the model training process. 

\begin{deluxetable}{c c l l}
    \tablecaption{The 2D-CNN architecture used for the calculation of core angular rotational speed, $\omega_0$.
    \label{tab_cnn1}}
    \tablehead{\colhead{Layer} & \colhead{Type} & \colhead{Output Shape} & \colhead{Param \#}}
    
    \startdata
    1  & Conv2D           & (None, 256, 256, 64) & 1664\\
    2  & MaxPooling2D & (None, 128, 128, 64) & 0 \\
    3  & Conv2D           & (None, 128, 128, 128) & 204928\\
    4  & MaxPooling2D & (None, 64, 64, 128) & 0 \\
    5  & Conv2D           & (None, 64, 64, 256) & 819456\\
    6  & MaxPooling2D & (None, 32, 32, 256) & 0 \\
    7  & Conv2D           & (None, 32, 32, 256) & 590080\\
    8  & MaxPooling2D & (None, 16,16, 256) & 0 \\
    9  & Conv2D           & (None, 16,16, 256) & 590080\\
    10 & MaxPooling2D & (None, 8, 8, 256) & 0 \\
    11 & Dropout & (None, 8, 8, 256) & 0 \\
    & & & \\
    \hline
    \colhead{Layer} & \colhead{Type} & \colhead{\# of Output Neurons} & \colhead{Param \#}\\
    12 & Flatten & (None, 16384) & 0\\
    13 & Dense & (None, 1024) & 16778240\\
    14 & Dense & (None, 512) & 524800\\
    15 & Dense & (None, 128) & 65664\\
    16 & Dense & (None, 32) &  4128\\
    \enddata
    

    \end{deluxetable}
\section{Re-examination of EoS classification}
\label{Appendix:EoS}

Although the results of Table \ref{Table:Clustering re-examination I} is an independent justification of results shown in Figure~\ref{Fig:Confusion matrix of EoS}, it cannot exclude the possibility to have subgroups within each group of Type I. In order to further investigate such possibility, we further re-examine above clustering type by another method: We select the data of each individual EoS to be the test data and all other data (calculatedf from other EoSs) to be the training data. We then repeat the model training and test by using different EoS as the test data. We could then obtain a series of accuracy provided by these models, which should provide how "close" of the test data are from the other training data.

In Table \ref{Table:Clustering re-examination II}, we show the calculated model accuracy for such a re-examination. 
The results of good binary classification shows that the EoS in the test data is somehow "close" to at least one EoS in the training data, so that the test data could be well-learned by the training data. On the other hand, the poor classification results indicates that the EoS in the test data could be considered "well-separated" from all other EoSs. According to the Table \ref{Table:Clustering re-examination II}, we could see that each EoS is close to at least one of others, but LS375 may be the only one, which could be considered "far" from others. Therefore, in principle, we may consider to separate LS375 as an independent group. But for the convenience of discussion, we will still keep it into the same group as LS180 and LS220 in this paper.

\begin{table}
\centering
\begin{tabular}{c c c c }
\hline\hline
Model & Test data & Training data & Accuracy \\
\hline
1 & LS180 & all non-LS180 & 90\% \\
2 & LS220 & all  non-LS220 & 97\% \\
3 & LS375 & all non-LS375 & 77\% \\
4 & GShenFSU1.7 & all non-GShenFSU1.7 & 100\% \\
5 & GShenFSU2.1 & all non-GShenFSU2.1 & 100\% \\
6 & GShenNL3 & all non-GShenNL3 & 100\% \\
7 & HShen & all non-HShen & 94\% \\
8 & HShenH & all non-HShenH & 89\% \\
9 & BHBL & all non-BHBL & 100\% \\
10 & BHBLP & all non-BHBLP & 100\% \\
11 & SFHo & all non-SFHo & 100\% \\
12 & SFHx & all non-SFHx & 100\% \\
13 & HSDD2 & all non-HSDD2 & 100\% \\
14 & HDFSG & all non-HDFSG & 100\% \\
15 & HSIUF & all non-HSIUF & 100\% \\
16 & HSNL3 & all non-HSNL3 & 100\% \\
17 & HSTM1 & all non-HSTM1 & 100\% \\
18 & HSTMA & all non-HSTMA & 100\% \\
\hline
\hline
\end{tabular}
\caption{Model accuracy by using a given EoS to be the test data and other EoS to be the training data (see the text). 
}
\label{Table:Clustering re-examination II}
\end{table}

  
\end{document}